\documentclass[twocolumn,pre,superscriptaddress,floatfix]{revtex4-1}
\pdfoutput=1  
\usepackage{amsmath}
\usepackage{amssymb}
\usepackage{amsfonts}   
\usepackage{graphics}   
\usepackage{graphicx} 
\usepackage{epsfig}    
\usepackage{rotating}  
\usepackage[latin1]{inputenc}
\usepackage{color}
\usepackage[caption=false]{subfig}

\begin{document}

\title{Language change in a multiple group society}

\author{Cristina-Maria Pop and Erwin Frey}

\affiliation{Arnold Sommerfeld Center for Theoretical Physics and Center for NanoScience, Department of Physics, Ludwig-Maximilians-Universit\"at M\"unchen, Theresienstrasse 37, 80333 M\"unchen, Germany}

%\date{\today}

\begin{abstract}
The processes leading to change in languages are manifold. In order to reduce ambiguity in the transmission of information, agreement on a set of conventions for recurring problems is favored. In addition to that, speakers tend to use particular linguistic variants associated with the social groups they identify with. The influence of other groups propagating across the speech community as new variant forms sustains the competition between linguistic variants. With the utterance selection model, an evolutionary description of language change, Baxter \textit{et al.\ }\cite{Baxter:2006p117} have provided a mathematical formulation of the interactions inside a group of speakers, exploring the mechanisms that lead to or inhibit the fixation of linguistic variants. 
In this paper, we take the utterance selection model one step further by describing a speech community consisting of multiple interacting groups. Tuning the interaction strength between groups allows us to gain deeper understanding about the way in which linguistic variants propagate and how their distribution depends on the group partitioning. Both for the group size and the number of groups we find scaling behaviors with two asymptotic regimes. If groups are strongly connected, the dynamics is that of the standard utterance selection model, whereas if their coupling is weak, the magnitude of the latter along with the system size governs the way consensus is reached. Furthermore, we find that a high influence of the interlocutor on a speaker's utterances can act as a counterweight to group segregation.
\end{abstract}

\maketitle

\section{Introduction}
\label{intro}

Language is one of the most prominent examples of complex systems. The phenomena underlying its emergence and evolution, as well as cultural change, have been subject to increased interest from the physics community over the past two decades. An account of the insight gained from the study of various mathematical models can be found in Refs.\ \cite{Baxter:2006p117,Castellano:2009,Sole:2010p1531,Pagel:2009p1443,Schulze:2008p831,Wang:2005p139,Nowak:2002p1441,Cangelosi:2002}. Confining our attention to the ongoing processes acting on vocabulary, we only need to monitor our own use of language as it changes over time to see that words emerge, modify their shape and disappear, and sometimes even alter their meaning or functionality. The reasons for these changes are diverse. In order to transmit a message as efficient and unambiguous as possible, a speaker will tend to use the conventions of her language. (From now on we will use the convention that when referring to the speaker, the female pronoun is used, whereas for the hearer we employ the masculine.)  However, since language does not only communicate meaning, but also reflects the speakers' cultural and social background, different linguistic variants (``different ways of saying the same thing'' \cite{Baxter:2006p117}) are used depending on the situation in which the speakers find themselves \cite{Croft:2000}. Also, because language ``needs to keep pace with new realities, new technologies and new ideas, from ploughs to laser printers, and from political-correctness to sms-texting'' \cite{Deutscher:2005}, new means of expressing an idea can enter the language of a community via innovations from its members. Thus, linguistic variants enter a competition for speakers \cite{Mufwene:2002}. It is this aspect in the change of languages that we have a closer look at in this paper.  

If speakers coming from distinct backgrounds find themselves united in a group, over time they will develop a common vocabulary in order to communicate successfully \cite{Steels:1995,Baxter:2009p834}.  As our society consists of many groups, defined either by geographical location, age, profession or other criteria, we notice two antithetic tendencies that dominate the dynamics of the language: on the one hand, speakers affiliated to a social group will try to reach consensus on a variant in order to describe a particular situation. On the other hand, since this variant can differ from group to group, an element of rivalry between various forms stems from the interactions between speakers belonging to distinct social groups \cite{Croft:2000}. Our aim is to understand how the competition among word variants is resolved in such a society composed of several groups of speakers connected with each other. To this end we investigate how long it takes, on average, until only one variant is being used throughout the speech community, and which conditions have to be met for consensus to be a realistic outcome. 

In the following, the term ``language'' is used to describe spoken language. It differs from written language insofar as it changes on a faster time scale, perceivable in the course of a human lifetime. The records of written language often do not capture the whole spectrum of changes in a language over shorter periods \cite{Sole:2010p1531,Graddol:2004p1459,New:2007p1534}, the reason being that some of them are alterations in pronunciation, which take a long time until they become reflected in the way the words are written, and others are short-lived vogue words. 

This paper is organized as follows: we first revisit the utterance selection model of language change in a group of speakers introduced by Baxter \textit{et al.\ }\cite{Baxter:2006p117}. Then we study the formation of consensus in a system composed of two coupled groups. With the aid of stochastic simulations, we find a scaling law for the fixation time of a variant and discuss its two asymptotic regimes. We characterize the dependence of the boundary between the asymptotic regimes on the parameters of the model. Finally, we generalize to systems of many coupled groups on two types of networks and again obtain scaling laws that we explain using analogies with well-known problems of statistical mechanics.

\section{Utterance Selection Model}
\label{sec:1}

Baxter \textit{et al.\ }\cite{Baxter:2006p117} formulated an evolutionary model of language change, termed utterance selection model of language change, based on the ideas presented by Croft \cite{Croft:2000}. It describes a very simple language, consisting of only one lingueme, i.e.,\ one object or situation to be described, and $V$ variants that can be used in referring to it. The underlying social network consists of $N$ speakers who use these variants according to their knowledge of language, with certain frequencies, which are stored in the vectors $\vec{x}_i=(x_{i1},x_{i2},...,x_{iV})$, with $i=1,...,N$. These frequencies are normalized to 1 for each speaker:

\begin{equation}
	\sum_{v=1}^V x_{iv}(t)=1, \ \ \forall i,t.
\end{equation}
This probabilistic representation of vocabulary can be encountered in various other models of language evolution \cite{Hurford:1989p72,Nowak:1999p68,Ke:2002p134}. The state of the population can be given at any given time $t$ through the entirety of agents' vocabularies $X(v)=(\vec{x}_1(t),...,\vec{x}_N(t))$. The speakers are placed on the nodes of a graph, and the probability of two of them interacting is given by the weight of the link connecting them.

The interaction algorithm between individuals relies on the following steps: first, two speakers are chosen at random, with the condition that they are connected by a link. In the next step, each of the speakers produces a string of tokens of length $T$. The tokens are instances of the $V$ variants, uttered according to frequencies $x'_{iv}$. If token production is unbiased, these frequencies are equal to the entries in the speaker's vocabulary vector, $x'_{iv}=x_{iv}$. If there is a bias in production, the probabilities to utter tokens of a particular variant are a linear transformation of these: $x'_{iv}=\sum_w M_{wv} x_{iw}(t)$, 
where the matrix $M$, same for all speakers, can be seen as the effect of universal forces such as articulatory constraints, according to Ref.\ \cite{Baxter:2006p117}. The columns of $M$ sum up to 1 so that the production frequencies are properly normalized. One of the effects of bias is that the speaker can also produce tokens of a variant that in her vocabulary has frequency zero, thus introducing innovations into the language. After both speakers have uttered their tokens, their vocabularies are updated, taking into consideration the old entries of the frequency vectors, the utterances of the speaker herself, as well as those of her interlocutor:

\begin{equation}
	\vec{x}_i(t+\delta t)=\frac{\vec{x}_i(t)+(\lambda/T)[\vec{n}_i(t)+H_{ij} \vec{n}_j(t)]}{1+\lambda(1+H_{ij})}.
	\label{update}
\end{equation}
In this expression, $\vec{n}_i$ and $\vec{n}_j$ are the number of tokens of each variant that have been uttered by speakers $i$ and $j$, respectively, so that $\sum_{v=1}^V n_{iv}=T$.

The parameter $\lambda$ gives the pressure for change exerted by an interaction on the vocabulary of the speaker. Since a speaker's vocabulary does not undergo dramatic changes in the course of an interaction, the value of $\lambda$ is generally taken to be small. 

The parameter $H_{ij}$ stands for the weight given by a speaker $i$ to her interlocutor $j$'s utterances relative to her own. Thus, for $H_{ij}$ smaller than 1, the interlocutor has a lower status than the speaker herself, whereas values of $H_{ij}$ larger than 1 would indicate a high status ascribed to the interlocutor by the speaker. In the update rule Eq.\ (\ref{update}), the denominator ensures the proper normalization of the frequencies. The steps of the algorithm are repeated until either there is only one variant spoken in the community (for unbiased token production) or another stationary distribution is reached (for biased production). 

If the variant production is unbiased, with time all but one variant will disappear, meaning that speakers will eventually reach consensus. How this is reached and how long it takes depends on the parameters of the model. If $H_{ij}$ is small, the speaker will mostly influence herself and thus, if she has a preferred variant, use this even more often, so that for most of the time each speaker will favor one variant over the others, but this preference can change due to interactions with other agents. If a variant has frequency zero in a speaker's vocabulary, it means it has fallen into disuse, so this speaker will never utter it again. For large values of $H_{ij}$, speakers have great influence on each other, and variants will spread across the community. The system converges toward a quasi-stationary distribution, from where a fluctuation will eventually drive it to consensus.

If tokens are produced with bias, the situation changes: the bias toward a particular variant prevents it from going extinct, but the frequency distribution can be such that either one variant is more common than the others, or several variants are used with more or less equal frequency \cite{Baxter:2006p117}.

In the following, we will concentrate on unbiased token production and study a simplified version of this model: the links between speakers have equal weight (the pairs of speakers will interact with the same frequency), all speakers ascribe the same weight to their interlocutor's utterances ($H_{ij} \equiv h$), and we restrict the number of variants to $V=2$. From Baxter \textit{et al.\ }\cite{Baxter:2006p117} we learn that this already allows the study of the most relevant types of behavior. They find that the time for the extinction of one variant (and thus fixation of the other) in a group of speakers is proportional to the system size squared, $t_c \propto N^2$. In Ref.\ \cite{Baxter:2008p82}, they show that the time to consensus is asymptotically network-independent. In the limit $h \to\infty$, one can consider an asymmetric version of the utterance selection model where in an interaction each agent behaves either as a speaker or as a listener. The case where the speaker is endowed with the ability to invent new variants and produces one token only per interaction is a minimal version \cite{Baronchelli:2006p116} of the model known as the ``naming game'' \cite{Steels:2005,Lenaerts:2005p135}. In the two-variant version, the dynamics corresponds to the voter model \cite{Blythe:2009,Krapivsky:2010}, one of the simplest models of opinion formation \cite{Balbas:2011}. In contrast, the symmetric utterance selection model, which we are considering in this paper, exhibits a much richer dynamics stemming from each agent being both speaker and listener at the same time.

\section{Multiple Group Utterance Selection Model}
\label{sec:2}

The utterance selection model gives a good insight into the linguistic dynamics of a group. However, society consists of many groups, with relatively weak connections among them. Thus, to better understand the mechanisms that cause languages to change at word level, we will examine this model in a wider context made up of several interacting non-overlapping groups. Blythe \cite{Blythe:2006p1516} studied the fixation probability and time for a variant in a subdivided population for two different spatial arrangements. For a system in which the groups are well-mixed, the time to consensus is proportional to the number of groups squared. If the groups are placed on a hub-and-spoke network, where in an interaction between groups one of these must be the hub, the fixation time approaches a constant, even in the limit of infinitely many groups. This is because, although the variants of the hub spread much faster, the large number of spokes ensures a finite fixation probability of a variant from one of these groups. The question concerning the influence of community structure on the emergence of consensus has also been addressed in the context of other models of language dynamics and opinion formation with two competing variants like the naming game \cite{Steels:1999,DallAsta:2006}, the voter, and the so-called AB model \cite{Castello:2006p140,Castello:2007}. In our approach, we analyze in detail the influence of increasingly strong separation of the groups on the time that a variant takes to fixate in the whole speech community. 

In this context, we introduce a new parameter, $f$, representing the ``group affinity'', that is, the probability of a speaker choosing her interlocutor from the same group. $1-f$ is then the probability that the speaker chooses a conversation partner from another group. Sood \textit{et al.\ }\cite{Sood:2008} studied this kind of varying coupling strength in the context of the voter model by allowing each agent belonging to a fully connected ``clique'' to be connected by an adjustable number of random links to agents belonging to a second clique. A similar parameter was employed by Baronchelli \textit{et al.\ }\cite{Baronchelli:2012} to describe the degree of mixing of two speech communities as the readiness of individuals belonging to one community to learn the language of the other community instead of their own.

Having restricted the number of variants to two, we can define a measure of consensus in a group, $x_0$, as the average over the first component of the frequency vector for all speakers (the frequency with which the first variant is used in the group), this being a number between zero and one. If $x_0$ is close to the ends of the interval, throughout that group one variant is used for most of the time. If, however, the value of $x_0$ is close to the center of the interval, speakers use both variants in significant proportions. This does not tell us, however, whether a speaker uses one preferred variant, which differs from speaker to speaker, or all use both variants with comparable frequencies. 

Regarding the initial conditions, we will fix half of the groups on one variant and the other half on the other variant. This way, the average time to consensus is larger than for uniformly distributed initial frequencies, because before global consensus on a particular variant can be reached, the variants  have to propagate across the groups. Random initial conditions would provide an already shuffled configuration, thus eliminating this mixing time. The time step between interactions, $\delta t$, is set to 1 for all simulations presented below. 

\subsection{Two groups}
\label{subsec:2:1}

The first step when moving from one group of speakers to a system composed of many groups is the coupling of two such entities.
\begin{quote}``Just imagine two groups living in two neighboring villages, speaking similar varieties of one language. With the passing of time, their language undergoes constant transformations, but as long as the two communities remain in close contact, their varieties will change in tandem: innovations in one village will soon spread to the other, because of the need to communicate. Now suppose that one of the groups wanders off in search of better land, and loses all contact with the speakers of the other village. The language of the two groups will then start wandering in different directions, because there will be nothing to maintain the changes in tandem''\cite{Deutscher:2005}.\end{quote}  Since the further away they are from each other, the less they interact, our parameter $f$ can be seen as a measure for the distance between the villages.

\begin{figure}[htb]	
	\centering 
	\includegraphics[width=0.9\columnwidth]{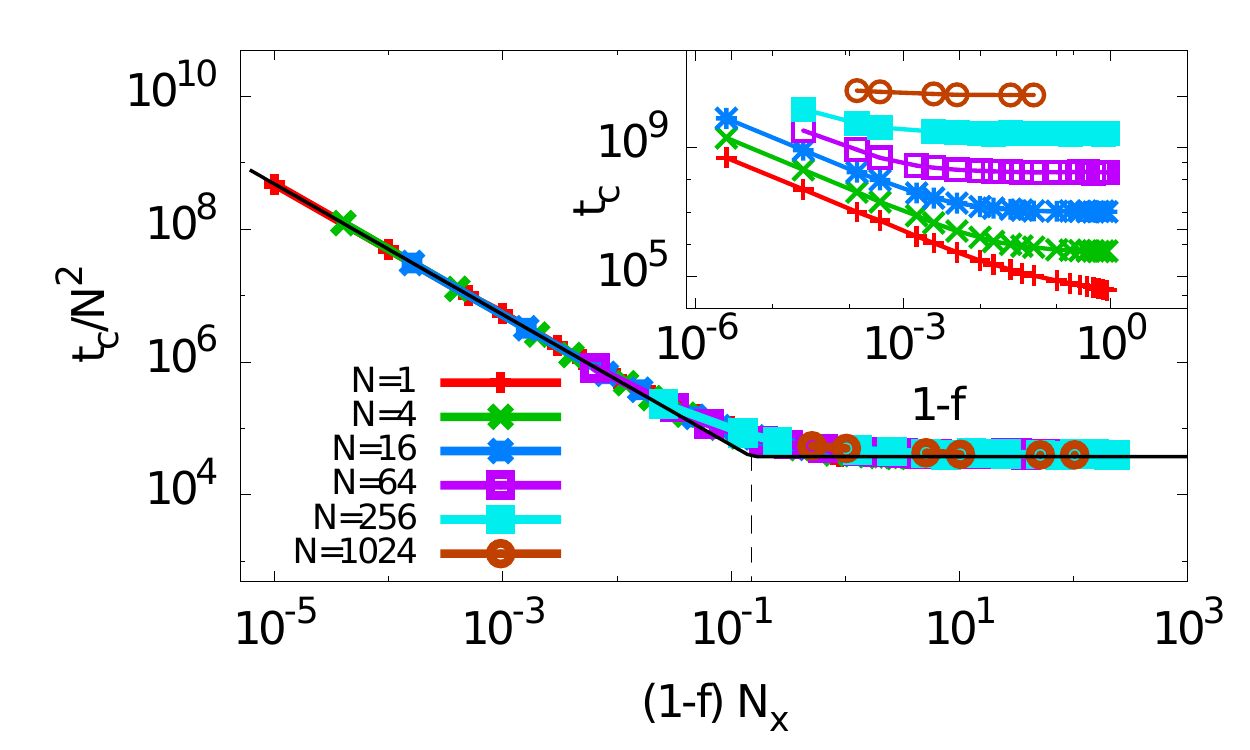}
	\caption{Scaling plot of the time to consensus as a function of group size $N$ and group affinity $f$ for two coupled groups. There are two asymptotic regimes, for strong and weak coupling respectively. The boundary is marked by the intersection $N_\times(1-f)$ of the curves fitting the power laws. The other parameter values are $T=1, \lambda=0.01, h=0.01$. Inset: The same curves before rescaling.}
	\label{scaleplot_inset}
\end{figure}

To understand the effect of the coupling, we impose the condition that the two groups are of the same size. 
As mentioned above, this scenario has been studied for the voter model update rule by Sood \textit{et al.\ }\cite{Sood:2008}. Varying the group size $N$, we numerically investigate the dependence of the average time to consensus on the group affinity parameter $f$ (Fig. \ref{scaleplot_inset}) and obtain scaling behavior: 
\begin{equation}
	t_c(f,N)=N^2 F\Big((1-f) N\Big).
	\label{scaling}
\end{equation}
Here the scaling function $F$ also depends on the number of tokens uttered by each speaker in an interaction, $T$, the pressure for change on the vocabulary of a speaker, $\lambda$, and the relative influence of the interlocutor, $h$.  

There are two asymptotic limits for the time to consensus, which are described by power laws.
For weak coupling between groups (large values of $f$), we find
\begin{equation}
	t_c(f,N)\propto N (1-f)^{-1}.
	\label{large_f}
\end{equation}
In contrast, if the coupling is strong, the function $F$ is constant, and therefore the average time to global consensus is given by
\begin{equation}
	t_c(f,N)\propto N^2.
	\label{small_f}
\end{equation}
The boundary between these two asymptotic regimes, $N_\times$, defined as the intersection of the above power laws, marks the transition from the one-group to the two-group behavior. While for group sizes $N>N_\times$, one is in the strong coupling regime, the two groups are only weakly coupled for $N<N_\times$. The reason for this is that for the same value of the group affinity $f$, a smaller group will restore inner consensus faster and thus its language will remain isolated from the one of the other group. In the case of larger groups, the new variant propagates more easily, such that the two groups share both variants for a longer time. 

We find that the value of the crossover group size, $N_\times$, depends on the parameters $T$, $\lambda$ and $h$ as follows:
\begin{equation}
(1-f) N_\times=\frac{1}{T}\eta(\lambda,h). 
\end{equation}

The result of $N_\times$ being inversely proportional to $T$, the number of tokens uttered, is in accordance with Baxter \textit{et al.\ }\cite{Baxter:2006p117}, where they find that $T$ enters the dynamics as a time scale. 

\begin{figure}[htb]
	\centering 
	\includegraphics[width=0.9\columnwidth]{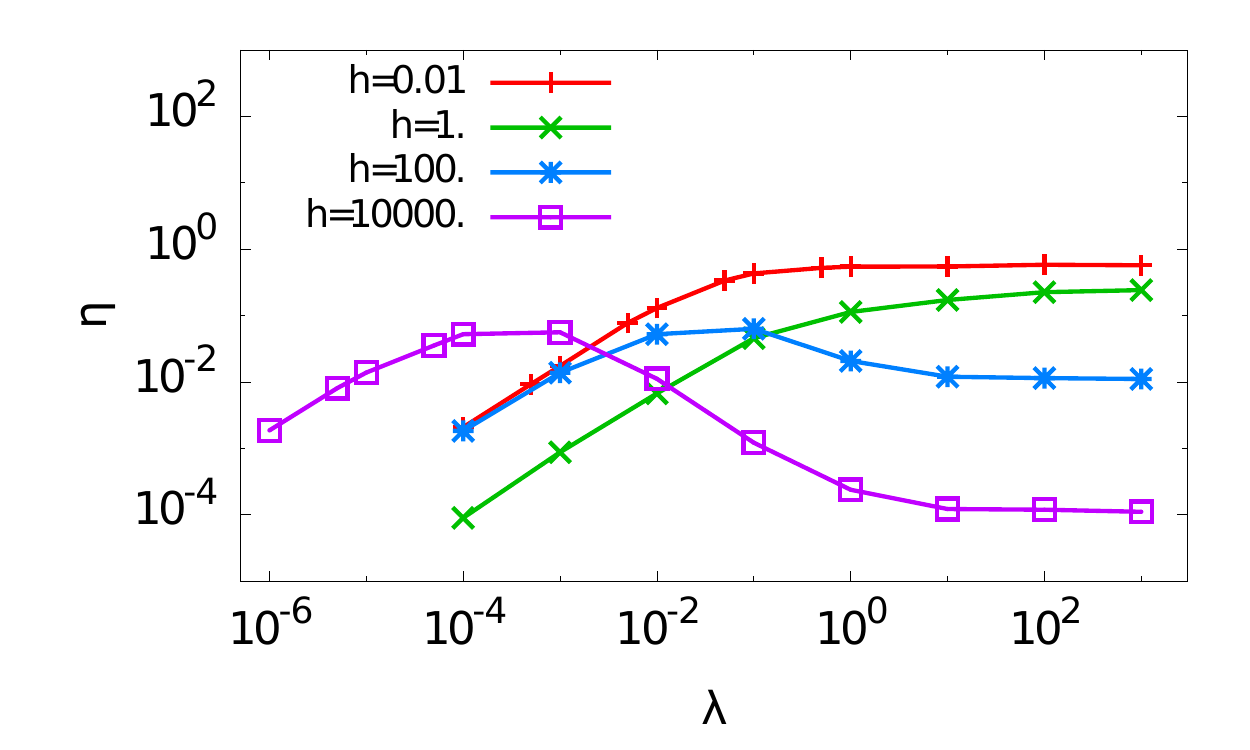}
	\caption{The dependence of the boundary between the asymptotic limits of the scaling law, $(1-f)N_\times$, on the magnitude of change in vocabulary $\lambda$ is linear for $\lambda<1/h$. For large values of $\lambda$, the function $\eta$ becomes independent of this parameter. Other parameter values are $N=2$ and $T=1$.}
	\label{depl}
\end{figure}

\begin{figure}[htb]
	\centering 
	\includegraphics[width=0.9\columnwidth]{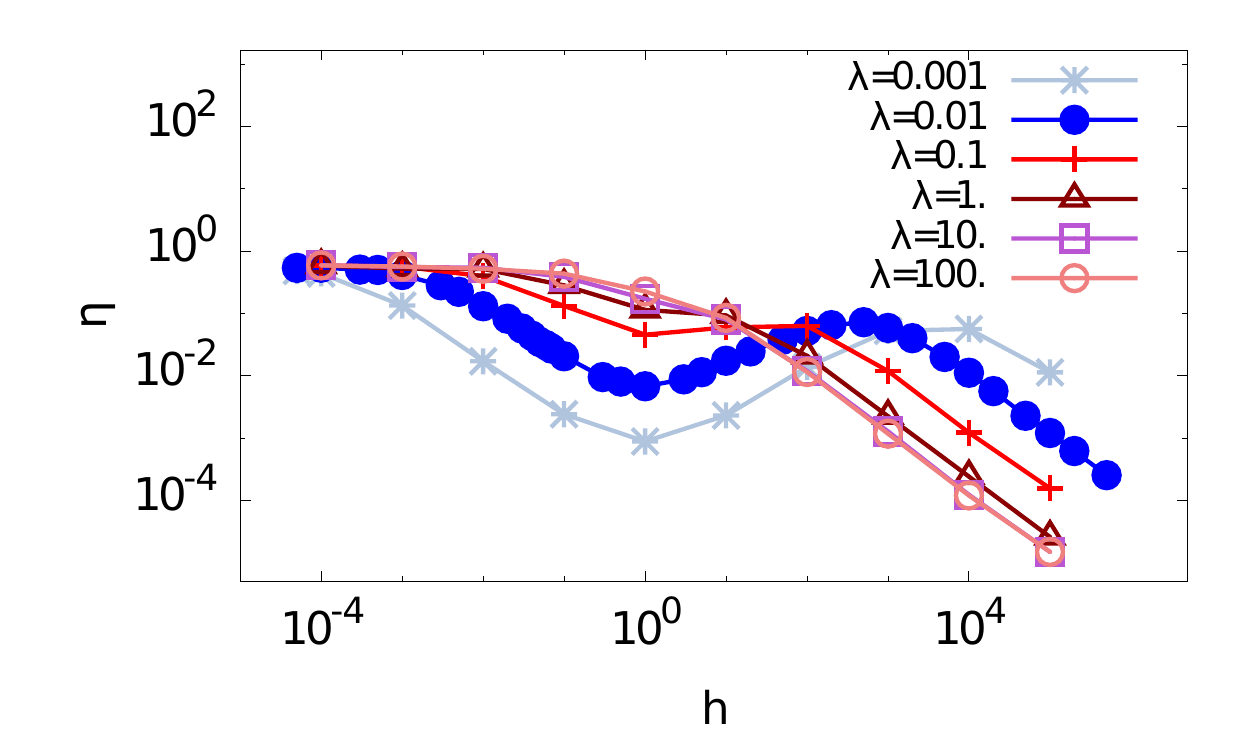}
	\caption{For very small values of the relative influence of the interlocutor, $h$, the boundary $(1-f)N_\times$ does not depend on this parameter. In the limit of large $h$, the boundary becomes proportional to $1/h$. The values of the other parameters are $N=2$ and $T=1$.}
	\label{deph}
\end{figure}

Due to the scaling behavior, Eq.~\eqref{scaling}, the location of the crossover point can either be found by varying $N$ for fixed $f$ or vice versa. We choose to keep the group size $N$ fixed and vary the coupling strength $f$ to determine $\eta(\lambda,h)$, which has a complex structure in terms of the parameters $\lambda$ and $h$, as can be seen in Figs. \ref{depl} and \ref{deph}. 

The parameter $\lambda$ sets the magnitude of change in vocabulary during an interaction. Investigating the dependence of the function $\eta$ on $\lambda$ with the aid of simulations for different values of $h$, we find that in the limit of small $\lambda$, $\eta$ depends linearly on this parameter. This reflects the fact that for $\lambda\to 0$, when interactions cannot change the vocabularies of the speakers any more, the number of groups makes no difference.  On the other hand, for very large $\lambda$, $\eta$ becomes constant, since the update rule Eq.\ \eqref{update} is now independent of $\lambda$, cf. Fig. \ref{depl}.

Regarding the role that the interlocutor's influence, $h$, plays in setting the value of the crossover, we can again characterize the limiting cases. For $h$ very small, one is in a non-interacting regime, where the utterances of the interlocutor hardly cause any changes in an agent's vocabulary. The speakers adopt the other variant very reluctantly, and the groups will only adjust their behavior toward each other if they are quite large, so that enough speakers from both groups have the chance to interact. In this limit, $\eta$ is independent of $h$, cf. Fig.~\ref{deph}.  In contrast, a large $h$ stands for a very strong influence of the interlocutor, so that the speaker chooses these utterances as her new vocabulary, and the old frequencies no longer play a role. This leads to speakers easily adopting the variant of an interlocutor from the other group and hereby to a strong coupling between groups. This however does not mean that consensus is reached faster, since the differences in vocabulary between speakers do not decrease significantly. In this last regime, the dependence of  $\eta$ on $h$ is a power law with exponent $-1$. This means that for $h\to\infty$, $N_\times$ approaches zero. Thus, the dynamics becomes independent of $f$, and the two groups will behave like one group. We see therefore that a large $h$ can counteract the effect of group segregation. A further observation is that for small $\lambda$, $\eta$ displays a distinct behavior around the point $h=1$. Approaching this point from below, the interlocutor's influence is large enough to allow the two agents involved in the interaction to align their vocabularies, and thus smaller group sizes are sufficient in ensuring a well-mixed behavior between the groups. At $h=1$, the utterances of the speaker and her interlocutor have the same weight, which enables rapid alignment of the speakers' vocabularies through the dissemination of the variants across the groups. For $h$ larger than $1$, the relative influence of the speaker's own utterances decreases, and the interlocutor's utterances dominate the interaction. The behavior of $\eta$ in this parameter range is symmetric with respect to $h=1$.

\begin{figure*}[htb]
	\centering%
	\resizebox{0.9\textwidth}{!}{
		\subfloat[]{\includegraphics{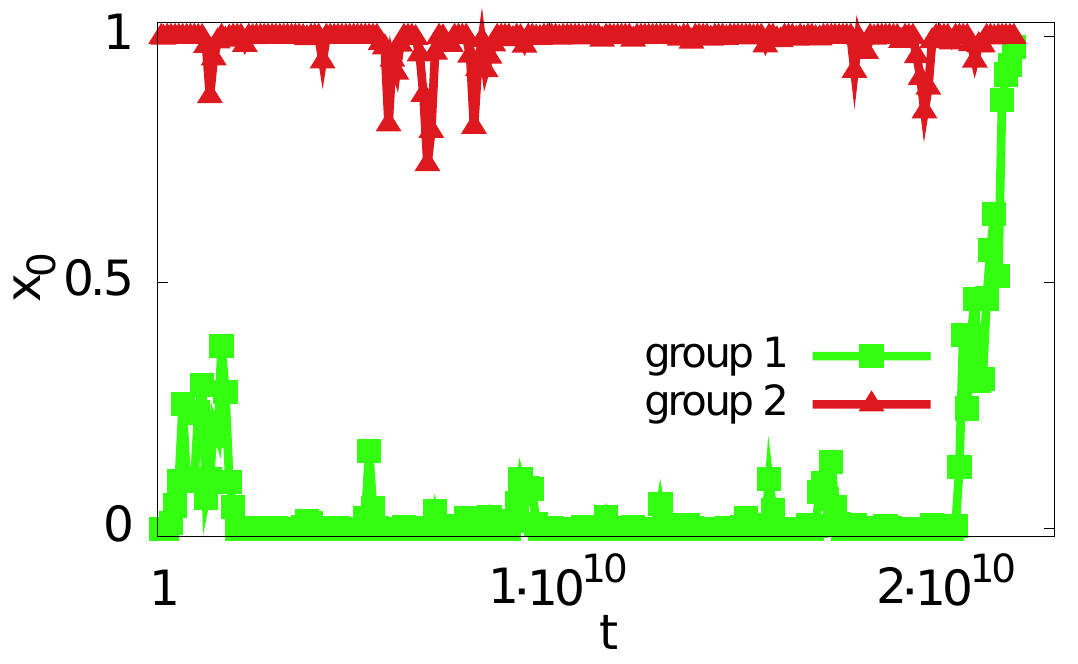}
			\label{trajs:largest}}
		\hfill
		\subfloat[]{\includegraphics{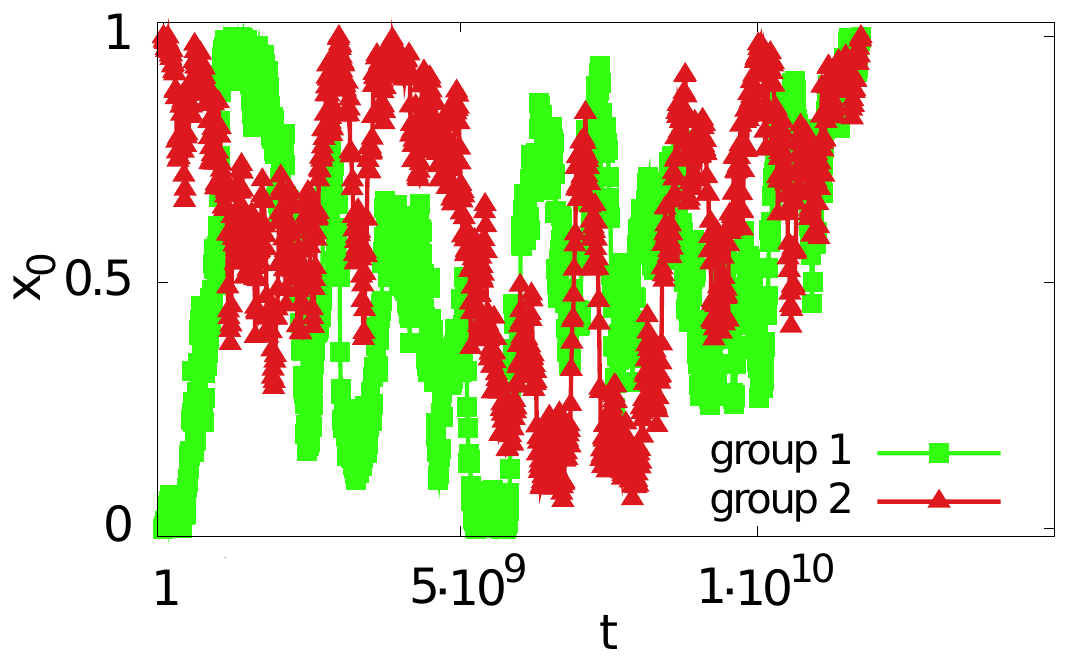}
			\label{trajs:large}}
		\hfill
		\subfloat[]{\includegraphics{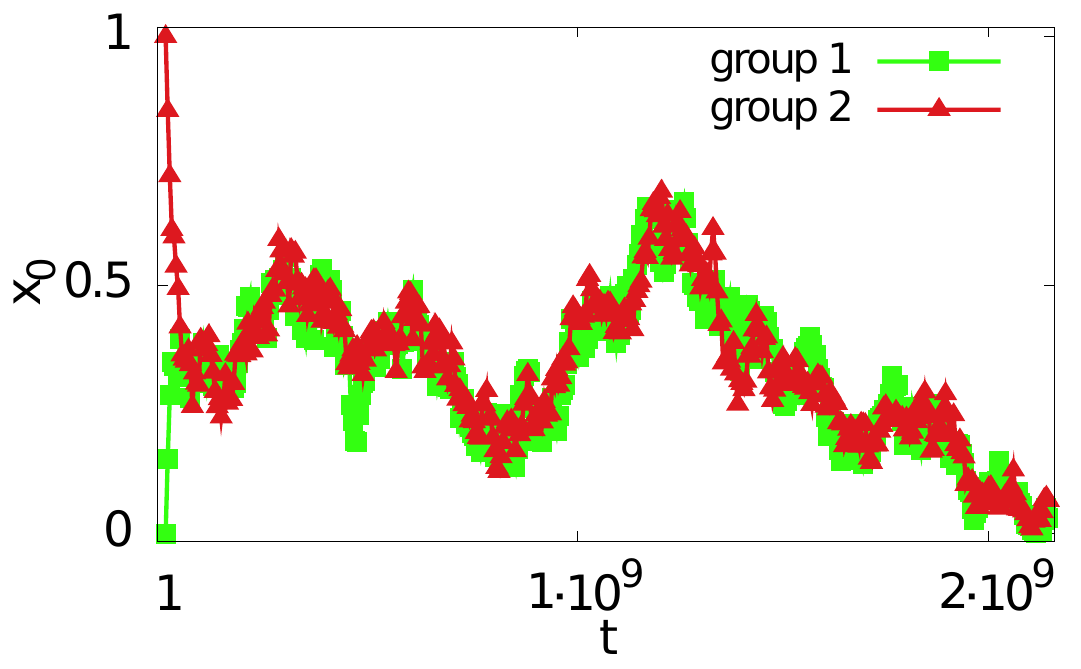}
			\label{trajs:small}}
		\hfill
		}
	\caption{Typical time trajectories of the consensus measure $x_0$ of two groups with \subref{trajs:largest} weak ($f=0.9999$), \subref{trajs:large} intermediate ($f=0.998$) and \subref{trajs:small} strong ($f=0.9$) coupling. Other parameter values: $T=1$, $\lambda=0.01$, $h=0.01$. }	
	\label{trajs}
\end{figure*}

To gain some understanding regarding the dynamics of the two groups in the various coupling regimes, we plot the consensus measure $x_0$ for each group (Fig. \ref{trajs}), considering three different values of the group affinity $f$ corresponding to weak, intermediate and strong coupling respectively. In Fig. \ref{trajs:largest} we see a trajectory where the groups ignore each other for most of the time. In Fig. \ref{trajs:large}, the more frequent interactions of the curves representing the measure of consensus in each group is the manifestation of a stronger coupling. Finally, in Fig. \ref{trajs:small}, the groups are coupled so strongly that they share the amount of consensus on a variant. 

In the following, we will discuss the weak and strong group coupling regimes in more detail.

\subsubsection{Weak coupling}
\label{subsubsec:2:1:1}

To understand Eq.\ (\ref{large_f}), remember that $f$ was the probability for a speaker to interact inside her own group. Then $1-f$ is the probability of an interaction with a speaker from the other group, and 
\begin{equation}
	\tau:\ =\displaystyle{\frac{1}{1-f}}
\end{equation}	 
is the average time between two interactions of this type. 
If $\tau$ is much larger than the average time to consensus in a group, the two groups will evolve independently, each of them reaching internal consensus, and perceive the interactions with the other group only as a series of perturbations (Fig. \ref{trajs:largest}). Eventually one of the perturbations leads one group to fixate on the variant spoken by the other group. The probability that the group will adopt the variant that the other group has agreed upon is $p=1/N$ (since in every interaction there are two speakers involved, in a conversation between groups one speaker out of $N$ is ``converted'' by her interlocutor  and then disseminates the opinion in her own group). The dynamics we are dealing with here is the well-known ``gambler's ruin'' problem \cite{Grinstead:2003}.  

If $\tau$ is the average time between inter-group interactions, the number of such interactions until consensus is reached at time $t$ is $$n \approx \displaystyle{ \frac{t}{\tau} = t\ (1-f)}.$$
Out of $n$ trials, the last one is successful, so the probability that the \textit{n}-th perturbation will lead to global consensus is $$P(n) = \Big(1-\frac{1}{N}\Big)^{n-1} \frac{1}{N}.$$ 
Since $P(n)$ are the terms of a geometric progression, the probability is properly normalized. The average time to consensus is then given by
\begin{equation}
	t_c = \tau \sum_{n=1}^\infty n P(n) \nonumber = \frac{N}{1-f},
\end{equation}	
which corresponds to the left part of the scaling law in Fig. \ref{scaleplot_inset}.

\subsubsection{Strong coupling}
\label{subsubsec:2:1:2}

For small values of $f$, the coupling between the groups is so strong that the groups will evolve towards each other, only to start diffusing together after reaching a common value (Fig. \ref{trajs:small}).  The two groups thus turn into one large group, a behavior already described by Baxter \textit{et al.\ }\cite{Baxter:2006p117}, with the time to consensus proportional to the system size squared:
\begin{equation}
	t_c \propto N^2.
\end{equation}

\begin{figure}[htb]
	\centering
	\resizebox{0.9\columnwidth}{!}{
		\subfloat[]{\includegraphics{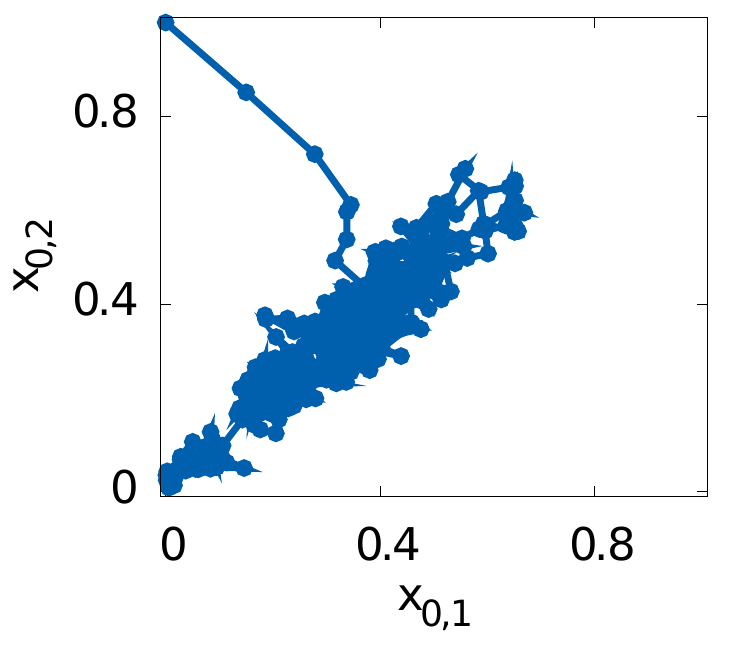}
			\label{x0square:fsmall}}
		\hfill	
		\subfloat[]{\includegraphics{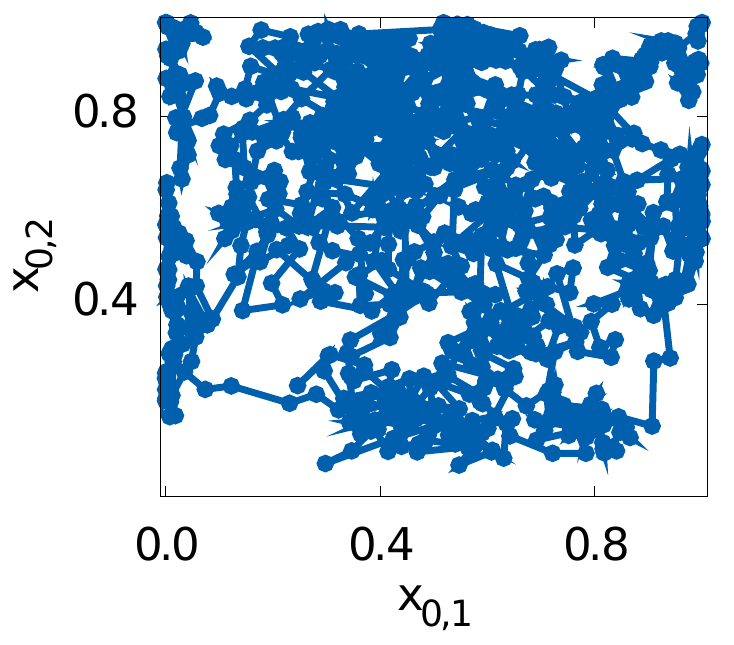}
			\label{x0square:finterm}}
			}
	\caption{ \subref{x0square:fsmall} The trajectory from Fig. \ref{trajs:small}, for strong group coupling, here in ($x_{0,1}$,$x_{0,2}$) coordinates. The dynamics is that of a one-dimensional random walk with absorbing boundaries. \subref{x0square:finterm} The trajectory for intermediate group coupling from Fig.~\ref{trajs:large}. It is the dynamics of a two-dimensional random walk in a rectangular area with reflecting boundaries and two absorbing points.}
	\label{x0square}
\end{figure}

Plotting the parameter measuring consensus for each of the two groups as a trajectory in $(x_{0,1},x_{0,2})$ coordinates (Fig. \ref{x0square:fsmall}), the dynamics is that of a biased random walk along the main diagonal of the square, which is quasi one-dimensional. Since there are only two absorbing points, namely $(0,0)$ and $(1,1)$, with increasing system size the ``escape windows'' become smaller and the time for one of them to be reached (which is the condition for consensus) diverges. This is commonly known as a narrow escape problem \cite{Singer:2008}.

With increasing $f$, this random walk frays more and more, to the point where it fills out the entire square (Fig. \ref{x0square:finterm} displays the trajectory shown in Fig. \ref{trajs:large}). This particular two-dimensional narrow escape problem was solved by Singer \textit{et al.\ }\cite{Singer:2006}. The mean first passage time shows a logarithmic correction in $N$, with respect to the previous result:

\begin{equation}	
	t_c \propto N^2\log N.
\end{equation} 

Being at the crossover of the strong and weak coupling regimes, the range of $f$ for which this dynamics is observed is so narrow that $N^2$ and $N^2 \log N $ cannot be distinguished in our numerical data. The $N^2$-dependence however is in good agreement with the simulation results. 

\subsection{Many groups}
\label{subsec:2:2}

Going one step further, we now fix the number of speakers in a group and instead vary the number of interacting groups. Simulations result in a scaling plot, similar to the one for two coupled groups. However, here the underlying phenomena are different. The scaling function for the average time to consensus has the form

\begin{equation}
	t_c=N_G^2 \tilde F(1-f),
\end{equation}
where again the function $\tilde F$ depends also on the other parameters of the model.
In a system with all-to-all connections between groups, for strong coupling the one-group result is found again, for the system size $N\cdot N_G$:

\begin{equation}
	t_c \propto (N\cdot N_G)^2.
	\label{tc_wm_strong}
\end{equation}	 
 
If the groups are weakly coupled, in addition to the number of groups and the group size, the average time between inter-group interactions plays an important role:  

\begin{equation}
	t_c \propto N_G^2\cdot N\cdot \tau.
	\label{tc_g_wm}
\end{equation}
 
On a two-dimensional square lattice, strong coupling leads to the same results as in the well-mixed case, thus $t_c$ is given by Eq.~(\ref{tc_wm_strong}). For large values of the parameter $f$, corresponding to weak coupling between the groups, there is a logarithmic correction due to the spatial arrangement of the groups:

\begin{equation}
	t_c\propto (N_G^2 \cdot \log N_G) \cdot N \cdot \tau.	
	\label{tc_g_l}
\end{equation}  

In the following we will provide a more detailed description of the system's behavior for both the well-mixed and the two-dimensional lattice configurations.

\subsubsection{Well-mixed system}
\label{subsubsec:2:2:1}

The simplest instance of a system composed of many connected groups is obtained by placing the groups on the nodes of a complete graph, meaning that each group interacts with each other group with equal probability.

\begin{figure}[htb]
	\centering
	\includegraphics[width=0.9\columnwidth]{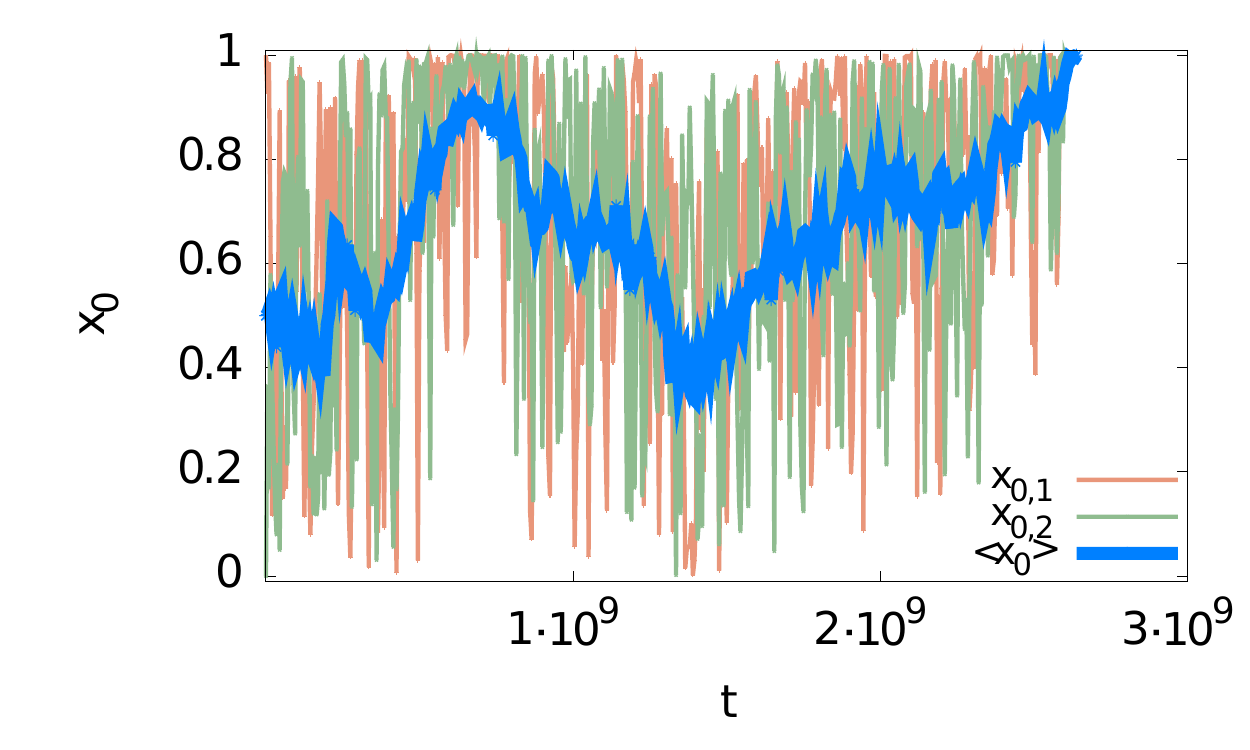}
	\caption{A typical coarse-grained trajectory of the consensus measure for two groups ($x_{0,1}$ and $x_{0,2}$, respectively), as well as the consensus measure throughout the system, $\langle x_0\rangle$, for $N_G=256$ groups with group affinity $f=0.5$ (other parameter values: $N=2$, $T=1$, $\lambda=0.01$, $h=0.01$). The points represent averages over 1000 simulation runs. For larger numbers $N$ of speakers in a group, the trajectory looks similar.}
	\label{Ng256_f05}
\end{figure}

If $f$ is small, i.e., the coupling between groups is strong, the two variants will diffuse across all the groups, both being used in each group for most of the time. In Fig. \ref{Ng256_f05} we see that the parameter measuring consensus for each group, $x_{0,i}$, takes values all over the interval $[0,1]$.  Again, we recover the time to consensus for one group, i.e.\ 
\begin{equation}
	t_c \propto (N\cdot N_G)^2.
\end{equation}	

\begin{figure}[htb]
	\centering
	\includegraphics[width=0.9\columnwidth]{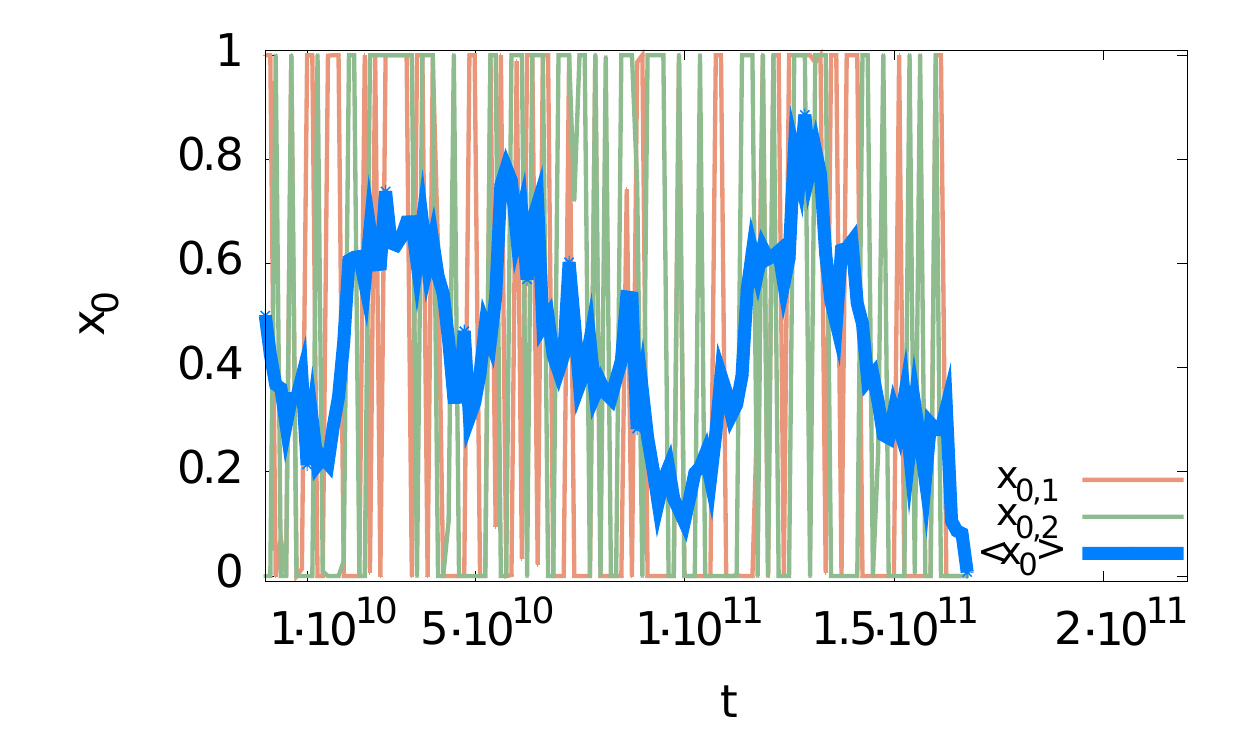}
	\caption{A coarse-grained trajectory of the consensus measures $x_{0,1}$, $x_{0,2}$ and $\langle x_0\rangle$ in a system of $N_G=256$ groups with group affinity $f=0.99$ (values of the other parameters: $N=2$, $T=1$, $\lambda=0.01$, $h=0.01$). The points are averaged over 1000 simulation runs.}
	\label{Ng256_f099}
\end{figure}

In the regime of large $f$, which denotes weak coupling, the average time for each of the two groups engaged in an interaction to achieve inner consensus is very short compared to the time scale $\tau$ of interactions between groups. Fig. \ref{Ng256_f099} shows that the parameters $x_{0,i}$ have either value 0 or 1 for most of the time, indicating the state of inner consensus. The parameter $\langle x_0\rangle$ however, which represents the measure of global consensus, fluctuates as a group changes its inner consensus from one variant to the other. When all groups speak the same variant, $\langle x_0\rangle$ reaches the value 0 or 1 and global consensus is achieved. 

When considering interactions between groups, there are three possible outcomes. If before the interaction, both groups shared consensus on the same variant, they remain in this state. If they were using different variants, after interacting, each of them can switch to the other variant with probability $p=1/N$ (as discussed already in the two-groups case). If both change to the respective other variant, globally it makes no difference, since the number of groups speaking each variant will be the same as before the interaction. If however only one of the groups changes to the variant of the other, the global balance is shifted towards one of the variants. This behavior corresponds to the voter model with link update \cite{Suchecki:2005p108,Castellano:2005,Sood:2008}, if time is rescaled so that on average one group opinion change takes place during every inter-group interaction. Here, instead of choosing a node and updating its opinion according to a randomly chosen neighbor, one chooses a link and updates the opinion of one of the two nodes involved. In networks with homogeneous degree distribution (here we have a complete graph), this choice has no effect on the results. The stochastic process is a one-dimensional random walk on the interval $[0,N]$ with absorbing boundaries \cite{Gardiner:2002}, which for our initial conditions (half of the groups starting with consensus on variant 0 and the other half on variant 1) results in a mean first passage time 
\begin{equation}
	t_c\propto N_G^2.\nonumber
\end{equation} 

On the much shorter time scale of a group changing opinion as a consequence of an interaction with its neighbor, the dynamics inside the group is again the one of the gambler's ruin problem, which we have encountered in the case of two groups with weak coupling. Summing up, the time to reach consensus in the setting of weakly connected groups on a complete graph is

\begin{equation}
	t_c \propto N_G^2\cdot N\cdot \tau.
	\label{tc_g_wm_gr}
\end{equation}

\begin{figure}[htb]
	\centering
		\includegraphics[width=0.9\columnwidth]{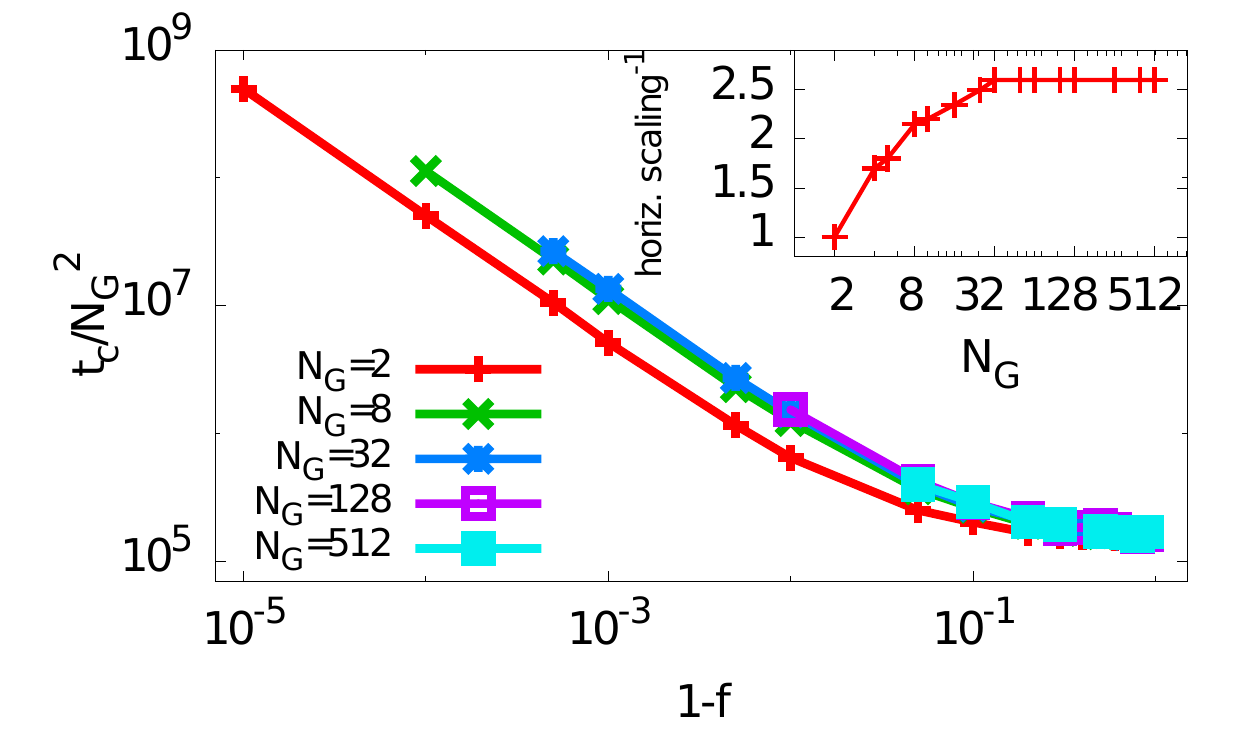}
	\caption{Scaling plot of the time to consensus for a well-mixed system with different number of groups and fixed group size. Inset: the horizontal scaling factor obtained by shifting the curves in order to obtain the master curve. For $N_G<32$, the scaling is different from the one for large $N_G$ due to finite-size effects. Other parameters: $N=4$, $T=1$, $\lambda=0.01$, $h=0.01$.}
	\label{scaleM_wm}
\end{figure}
In the scaling plot in Fig. \ref{scaleM_wm} we see that the left part of the curves for $N_G < 32$ does not overlap with the master curve, but is slightly parallel to it. This is due to finite size effects, since consensus for a small number of groups is reached in a somewhat different way than for large $N_G$. For $N_G=2$, global consensus is reached when one of the groups fixates on the variant of the other, so only one change of opinion is needed. If there are more than two groups involved, a group can change opinion several times before finally all line up, due to interactions with other groups using different variants. Simulations suggest that the boundary where the many-group behavior sets in is $N_G=32$ (see inset of Fig. \ref{scaleM_wm}).

\subsubsection {Groups on a lattice}
\label{subsubsec:2:2:2}

\begin{figure}[htb]
	\centering
		\includegraphics[width=0.9\columnwidth]{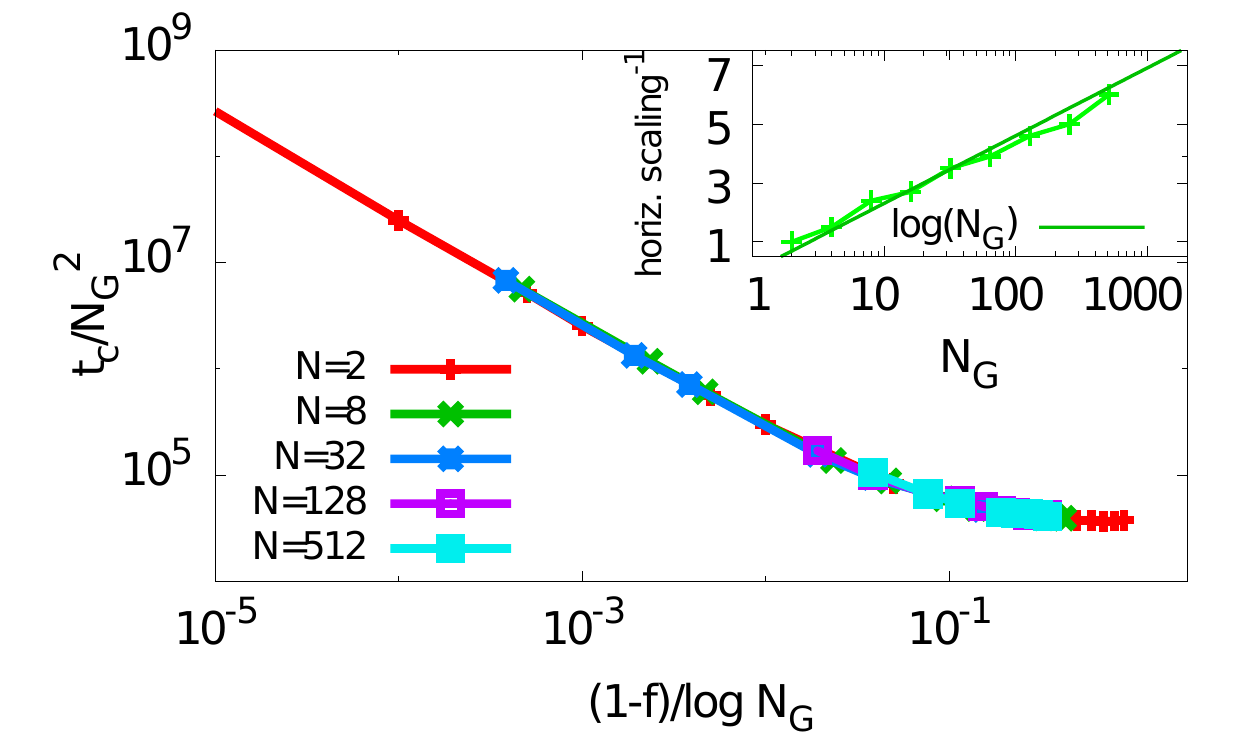}
	\caption{Scaling plot of the time to consensus for a system of groups on a square lattice. Inside the groups the configuration is well-mixed. Inset: horizontal scaling factor, obtained by shifting the simulation data curves so that they all fall onto the master curve. Other parameters: $N=2$, $T=1$, $\lambda=0.01$, $h=0.01$.}
	\label{M_scale_lat}
\end{figure}

We now place the groups on a two-dimensional square lattice, with each lattice site being occupied by exactly one group. Each group is thus allowed to interact with its four direct neighbors, and the boundary conditions are periodic. As in the well-mixed case, we obtain a scaling plot for the time to consensus (Fig. \ref{M_scale_lat}). 
For frequent interactions between groups, we have as before
\begin{equation}
	t_c\propto (N_G \cdot N)^2.
\end{equation}
The difference with respect to the well-mixed case is expressed through a $\log N_G$ correction. This is not surprising, since we are again looking at the voter model, this time on a two-dimensional lattice. For this, the time to consensus regardless of the initial conditions has been found to be $t_c \propto N_G \log N_G$ (\cite{Cox:1989,Krapivsky:2010}). Imposing the initial conditions that half of the groups start with consensus on one variant and the other half on the other variant, this turns into $t_c \propto N_G^2 \log N_G$ on the time scale $\tau$. Completing the picture with the time needed for a group to change its opinion, proportional to the group size $N$, the final result is
\begin{equation}
	t_c\propto (N_G^2 \cdot \log N_G) \cdot N \cdot \tau.	
	\label{tc_g_l_gr}
\end{equation} 

Thus, as expected, if groups display a spatial arrangement and interact only with their next neighbors, diversity of variants is preserved longer than in a system where all groups can interact with each other. This slower convergence time due to spatial constraints has also been found for the naming game and the AB model \cite{Baronchelli:2006p1561,Castello:2009}, with individual agents placed on the nodes of a lattice.

\section{Conclusions}
\label{concl}
In this work, we have extended the utterance selection model \cite{Baxter:2006p117} by giving the underlying social network a more complex structure, allowing for the existence of well-de\-li\-mi\-ted groups inside which speakers interact more often than with the rest of the speech community. We introduced the group affinity $f$, giving the probability that a speaker chooses his interaction partner to be from the same group, which we used for tuning the strength of the interactions between groups. Our object of interest, the average time until consensus is reached throughout the system, turns out to be highly sensitive to this parameter. Group structure is important, in that it gives rise to various types of behavior, depending on the size and the number of the groups, as well as the status of the interlocutor.

Upon investigating consensus formation in two interacting groups, we obtain a scaling law for the time needed until only one variant is used throughout the speech community. The results tell us that global consensus would be seriously impeded if the groups were too large or the interactions between them very scarce. The asymptotic limits of the scaling function show that for strong coupling the entire system behaves like one large group, and global consensus is reached in an average time proportional to the group size $N$ squared. If we further reduce the coupling strength, the average consensus time becomes proportional to the time interval between inter-group interactions, $\tau=1/(1-f)$, and the group size: $t_c \propto \tau N$. Global consensus is achieved when one of the groups switches to the variant used by the other group, the dynamics corresponding to the gambler's ruin problem. 

The boundary between the one-group and the many-group regime has a nontrivial dependence on the parameter $h$, which represents the influence that the interlocutor's utterances have on the vocabulary with respect to the speaker's own. When this parameter is very small, the speakers ignore each other almost completely, and the variants mix inside a group only if the latter is large enough. With increasing $h$, speakers start taking into account the utterances of their conversation partner, hereby contributing to the mixing of variants and thus decreasing the critical system size. For very large $h$, the speakers become ``amnesic'', meaning that their old vocabulary hardly plays a role any more and they orient their new vocabulary almost entirely after the utterances of the interlocutor. Due to the speakers being highly influenceable, variants spread across the whole system, which behaves like one large group in the limit $h\to\infty$. This means that a large $h$ can counteract even very weak coupling of the groups. 

For many coupled groups, a strong connection between them induces a single-group behavior, as all speakers start using both variants, thus again consensus is reached in a time proportional to the total number of speakers squared $(N\cdot N_G)^2$. If groups are more isolated, on the time scale of inter-group interactions the behavior is more complex. Even though groups might reach inner consensus on a variant, they might change their opinion several times, after interacting with other groups using different variants, before all agree on one variant and achieve global consensus. This is the dynamics of the voter model with link update for $N_G>2$, and is quite different from the two-group case, where it was enough for one group to change opinion once. As the number of groups is increased, we observe a finite size effect in the scaling factors. Again, the average time between two interactions between speakers belonging to different groups plays an important role. For a system where the groups are placed on a well-mixed network, the average time to consensus is $t_c \propto N_G^2\cdot N\cdot \tau$. In the weak coupling limit, the quadratic dependence on the number of groups is owed to the voter model dynamics. Inside a group, one speaker out of $N$ introduces a new variant as a result of an interaction with another group (which takes place on average every $\tau$ time steps). Since this dynamics is equivalent to the gambler's ruin problem, a $N\cdot \tau$ term arises. If the groups are positioned on the sites of a square lattice, a  logarithmic correction ensues due to the spatial arrangement, and $t_c\propto (N_G^2 \cdot \log N_G) \cdot N \cdot \tau$. In the same way as in the case of two groups, the parameter $h$ controls the position of the boundary between the asymptotic regimes in a complex manner.

We thus learn that not only strong segregation of the various groups, but also excessive partitioning of the speech community can lead to difficulties in reaching global consensus in a realistic period of time. However, ``insecure'' speakers, ascribing their interlocutors a much higher importance than themselves and adopting their vocabulary, can accelerate the establishing of a convention.

Latest technological developments are making records of spoken language more and more accessible: the Corpus of Contemporary American English (COCA) \cite{Davies:2007p1535} contains a large record of digitalized television and radio shows, and offers the tools to compare relative frequencies of words. Following a different approach, New \textit{et al.\ }\cite{New:2007p1534} have set up a data base of movie subtitles collected from the Internet and used it to approximate word frequencies in human interactions. Data collections of this type could offer valuable insights for future research related to the dynamics of linguistic variants.

\begin{acknowledgments}
We thank Richard Blythe and Gareth Baxter for discussions. Financial support from the Bavarian Academic Center for Central, Eastern and South Eastern Europe (BAYHOST) is gratefully acknowledged. 
\end{acknowledgments}

\end{document}